\documentclass[conference]{IEEEtran}
\IEEEoverridecommandlockouts

\usepackage{cite}
\usepackage{amsmath,amssymb,amsfonts}
\usepackage{algorithmic}
\usepackage{graphicx}
\usepackage{textcomp}
\usepackage{xcolor}
\usepackage{mdframed}
\usepackage{svg}
\usepackage{url}
\usepackage{booktabs}
\def\BibTeX{{\rm B\kern-.05em{\sc i\kern-.025em b}\kern-.08em
    T\kern-.1667em\lower.7ex\hbox{E}\kern-.125emX}}
\begin{document}

\title{Kumo: A Security-Focused Serverless Cloud Simulator
}


\author{
\IEEEauthorblockN{
    Wei Shao\IEEEauthorrefmark{1},
    Khaled N. Khasawneh\IEEEauthorrefmark{2},
    Setareh Rafatirad\IEEEauthorrefmark{1},
    Houman Homayoun\IEEEauthorrefmark{1}, and
    Chongzhou Fang\IEEEauthorrefmark{3}
}

\IEEEauthorblockA{\IEEEauthorrefmark{1}University of California, Davis, USA\\
Email: \{wayshao, srafatirad, hhomayoun\}@ucdavis.edu}
\IEEEauthorblockA{\IEEEauthorrefmark{2}George Mason University, USA\\
Email: kkhasawn@gmu.edu}
\IEEEauthorblockA{\IEEEauthorrefmark{3}Rochester Institute of Technology, USA\\
Email: cxfeec@rit.edu}
}

\maketitle

\begin{abstract}

Serverless computing abstracts infrastructure management but also obscures system-level behaviors that can introduce security risks. Prior work has shown that serverless platforms are vulnerable to attacks exploiting shared execution environments, including attacker--victim co-location and denial-of-service through resource contention, yet analyzing these risks on production platforms is difficult due to limited observability, high cost, and lack of experimental control, while existing simulators primarily focus on performance and cost rather than security. We present Kumo, a security-focused simulator for serverless platforms that enables controlled, reproducible analysis of security risks arising from scheduling and resource sharing decisions. Kumo models invocation arrivals, scheduler placement, container reuse, resource contention, and queuing within a discrete-event framework, explicitly representing attackers and victims as first-class entities and providing metrics such as co-location probability, time to first co-location, invocation drop rate, and tail latency. Through two case studies, we show that scheduler choice is a first-order factor for co-location attacks, inducing orders-of-magnitude differences under identical workloads, while Denial-of-Service behavior is largely governed by system-level factors such as service time, queuing policy, and cluster capacity once contention dominates. These results highlight the need to distinguish scheduler-driven isolation risks from broader resource exhaustion vulnerabilities and position Kumo as a flexible foundation for systematic, security-aware exploration of serverless platforms.

\end{abstract}

\begin{IEEEkeywords}
Serverless computing, cloud security, multi-tenant isolation, scheduling, resource contention, denial-of-service attacks, simulation, cloud platforms.
\end{IEEEkeywords}

\section{Introduction}
\label{sec:introduction}

Serverless computing is widely adopted for its elasticity, simplified deployment, and fine-grained billing. By abstracting away infrastructure management, however, serverless platforms also obscure scheduling, resource sharing, and container reuse decisions that can have important security implications~\cite{li2022serverless,shafiei2022serverless,marin2022serverless,ghorbian2024survey}.

Recent work shows that serverless platforms are vulnerable to attacks that exploit shared execution environments rather than software bugs~\cite{fang2022repttack,shao2025bit,zhao2024everywhere,nazari2023adversarial,fang2023gotcha,fang2025assessing}. Co-location attacks exploit placement decisions to place attacker and victim functions on the same worker, while availability attacks exploit contention and queuing to degrade victim performance or cause drops. Studying such attacks on production platforms is difficult due to limited observability, high cost, and poor controllability.



Existing serverless simulators and performance models primarily focus on cost estimation, latency optimization, or capacity planning. While valuable, these tools are not designed to analyze security risks, and typically lack explicit attacker modeling, isolation metrics, or the ability to study scheduler behavior as a security mechanism~\cite{mahmoudi2021simfaas,raith2023faas,mampage2023cloudsimsc}. As a result, there is currently no principled way to explore how serverless design choices influence security outcomes under controlled and reproducible conditions.

In this paper, we present \textbf{Kumo}, a security-focused simulator for serverless platforms. Rather than replicating any specific commercial platform in full detail, Kumo captures the mechanisms most relevant to serverless security, including scheduler placement, container reuse, resource contention, and queuing. It explicitly models attackers and victims and reports security-relevant metrics such as co-location probability, time to first co-location, invocation drop rate, and tail latency.

We demonstrate Kumo through two case studies. The first analyzes attacker--victim co-location under different scheduling policies and shows that scheduler choice can lead to orders-of-magnitude differences in co-location probability. The second examines Denial-of-Service (DoS) behavior and shows how availability degradation depends on service time, queuing policy, and cluster capacity. Together, these studies distinguish scheduler-driven isolation risks from broader resource exhaustion vulnerabilities.

This paper makes the following contributions:

\begin{itemize}
    \item We design and implement \textbf{Kumo}, a configurable, event-driven simulator for security analysis of serverless platforms, with explicit modeling of scheduling, container reuse, resource contention, and queuing.
    \item We introduce a flexible framework for modeling attackers and victims at the workload level, enabling direct measurement of security-relevant outcomes such as co-location probability, time to first co-location, and availability degradation.
    \item We present two in-depth case studies that demonstrate how scheduler design and system-level resource management influence distinct classes of serverless security risks.
\end{itemize}

By enabling controlled, reproducible exploration of serverless security tradeoffs, Kumo complements empirical studies on production systems and provides a foundation for principled analysis of serverless platform design.

\section{Background and Threat Model}
\label{sec:background}

This section provides brief background on the serverless execution model and defines the threat models considered in this work.

\subsection{Serverless Execution Model}

In serverless computing, applications are deployed as functions that are executed on demand in response to events. A cloud provider manages a pool of workers that host function containers and allocates resources dynamically. When a function is invoked, the platform selects a worker, initializes or reuses a container, and executes the function. Containers may be reused across invocations, leading to \emph{warm starts}, or freshly initialized, resulting in \emph{cold starts} that incur additional latency.

Serverless platforms are inherently multi-tenant: functions from different users may execute on the same worker and share underlying hardware and software resources. Placement decisions are typically handled by platform schedulers, which are opaque to tenants and optimized primarily for performance and resource efficiency. These characteristics make serverless platforms susceptible to security risks that arise from shared execution environments, even in the absence of software vulnerabilities.

\subsection{Threat Model}

We consider adversaries that exploit legitimate serverless execution mechanisms rather than implementation bugs. In all cases, attackers issue only valid function invocations and do not compromise the underlying platform or worker nodes.

\subsubsection{Co-location Attacks}

In a co-location attack, an adversary attempts to place its function execution on the same worker as a victim tenant via exploiting scheduling features~\cite{ishakian2010colocation,azar2014co,fang2023heteroscore}. Successful co-location enables a range of potential side-channel and information leakage attacks by exploiting shared resources~\cite{gruss2016flush+,kocher2020spectre,lipp2020meltdown,yarom2014flush+}. The attacker does not observe placement decisions directly, but can repeatedly invoke functions to increase the probability of sharing a worker with the victim. An attack is considered successful if attacker and victim functions are co-located on the same worker at any point during execution.

\subsubsection{Availability (Denial-of-Service) Attacks}

In an availability attack, an adversary seeks to degrade victim performance by inducing resource contention~\cite{nabi2016availability,sharifi2012availability}. The attacker injects a high volume of function invocations that compete for worker resources, leading to queuing delays, invocation drops, or increased tail latency for victim functions. These attacks do not overwhelm the network or exploit crashes; instead, they rely solely on the platform’s resource management and admission control behavior.

\subsection{Out-of-Scope Attacks}

Kumo does not model attacks that require low-level microarchitectural leakage, such as cache or speculative execution side channels, nor does it consider network-level distributed DoS attacks. This scope allows Kumo to focus on system-level security risks that emerge from shared execution environments and resource management decisions.

\section{Design}
\subsection{Design Goals and Scope}
\label{sec:design-goals}

Kumo is designed as a security-focused simulator for serverless platforms. Its primary goal is not to replicate any specific commercial platform in full detail, but to enable controlled, reproducible analysis of security risks that arise from cloud control plane decisions, including scheduling and resource sharing. To this end, Kumo is guided by the following design goals.

\paragraph{G1: Security-Oriented Modeling}
Kumo prioritizes modeling features that are essential for analyzing serverless security risks, including scheduler placement decisions, container reuse (cold vs.\ warm starts), resource contention, and queuing behavior. Rather than targeting microarchitectural accuracy, Kumo focuses on capturing the system-level mechanisms that enable attacks such as co-location and DoS.

\paragraph{G2: Explicit Attacker and Victim Modeling}
Unlike performance-oriented simulators, Kumo treats attackers and victims as explicit modeling primitive entities whose identities, behaviors, and outcomes are explicitly modeled. The simulator explicitly supports attacker-controlled workloads that interact with benign tenants through shared platform resources, enabling direct measurement of security-relevant outcomes such as co-location probability, time to first co-location, invocation drops, and tail latency.

\paragraph{G3: Reproducible and Controlled Experiments}
Kumo is designed to support reproducible experimentation through deterministic event-driven execution and configurable random seeds. Platform parameters, workloads, schedulers, and attacker behavior are all specified explicitly, allowing systematic exploration of security tradeoffs under controlled conditions.

\paragraph{G4: Scheduler Pluggability}
Scheduling policies are implemented through a clean abstraction that decouples scheduler logic from the execution engine. This allows researchers to evaluate existing serverless schedulers, prototype new policies, and study scheduler behavior as a security mechanism without modifying the core simulator.

\paragraph{G5: Low Barrier to Extension}
Kumo is intended to be easily extensible. New workloads, schedulers, attack strategies, and metrics can be added with minimal changes to the core engine, enabling rapid prototyping and iterative security analysis.


\subsection{Architecture Overview}
\label{sec:architecture}

\begin{figure}[htb!]
\centering
\includegraphics[width=.9\linewidth]{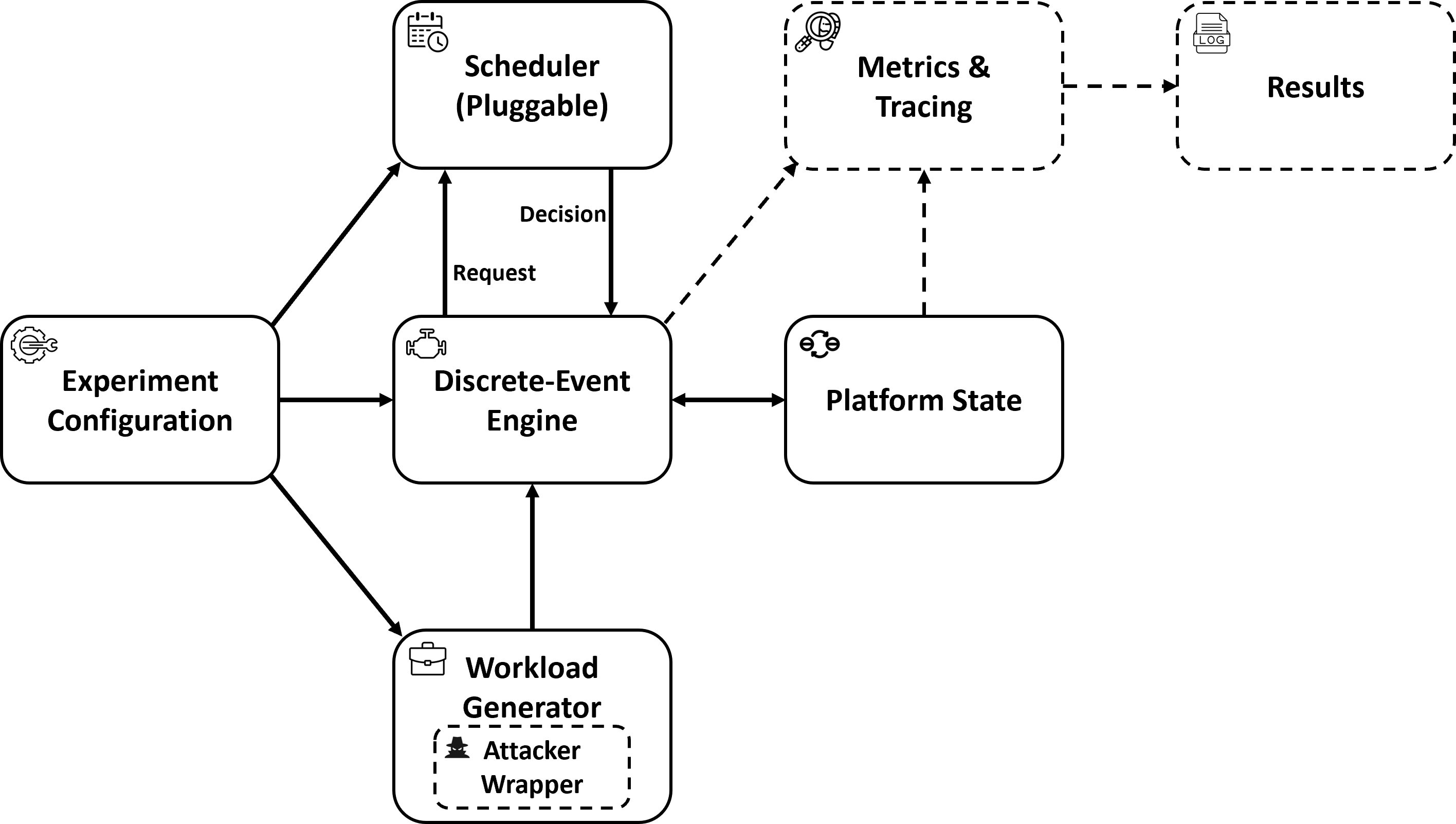}
\caption{Architecture of Kumo.}
\label{fig:kumo-architecture}
\end{figure}

Kumo is a discrete-event simulator that models serverless execution as a sequence of invocation arrivals, scheduling decisions, execution events, and resource reclamation. Its architecture cleanly separates workload generation, scheduling policy, platform state, and metric collection, enabling security-focused analysis under controlled and reproducible conditions. Figure~\ref{fig:kumo-architecture} provides an overview of Kumo’s architecture and execution flow.

At the core of Kumo is an event-driven simulation engine that advances logical time by processing a priority queue of events. Each function invocation is represented as a series of events, including invocation arrival, scheduling, execution start, and execution completion. This design allows Kumo to efficiently simulate large-scale workloads while preserving precise ordering of security-relevant events such as placement decisions, container reuse, and resource contention.

Invocation arrivals are produced by pluggable workload generators that emit batches of invocations over simulated time. Kumo supports multiple workload models, currently including uniform, Poisson, and bursty arrival processes. To enable security analysis, workloads may be wrapped by attacker modules that inject adversarial invocations alongside benign traffic. Generated invocations are timestamped and submitted to the engine without embedding any scheduling assumptions, ensuring that placement behavior is determined solely by the scheduler under study.

Scheduling decisions are handled through a dedicated scheduler abstraction. Upon invocation arrival, the engine queries the scheduler with the invocation metadata and the current platform state. The scheduler returns a placement decision indicating a selected worker or a failure if no placement is possible. By decoupling scheduling logic from execution, Kumo enables direct comparison of different scheduling policies under identical workloads and platform configurations, isolating the security implications of scheduler behavior.

Kumo maintains an explicit platform state that models workers, containers, and resource usage. Each worker tracks its available CPU, memory, and storage resources, as well as the set of active and idle containers. Containers are associated with specific functions and may be reused across invocations, capturing instance cold and warm start behavior. Resource allocation and reclamation are modeled at invocation start and completion, allowing contention, queuing, and capacity exhaustion effects to emerge naturally during simulation.

When worker resources are insufficient to immediately execute an invocation, Kumo may enqueue the invocation in a per-worker waiting queue, subject to a configurable queue length limit. If the queue is full, the invocation is dropped. This explicit modeling of backpressure is essential for analyzing availability degradation and DoS behavior under overload.

Security- and performance-relevant metrics are collected through a centralized metrics module that observes key events during simulation. Metrics include performance metrics like cold and warm start rates, invocation drop rates, end-to-end latency distributions, and security-related metrics like co-location counts and time to first co-location which are important to micro-architectural side-channel attackers~\cite{fang2022repttack,fang2023heteroscore,zhao2024everywhere}. Metrics collection is decoupled from the execution engine, ensuring that measurement does not influence scheduling or execution behavior.

In a typical execution, workloads generate invocation arrivals that are submitted to the engine. For each invocation, the engine queries the scheduler for placement, updates the platform state accordingly, and schedules execution and completion events. Throughout this process, the metrics module records security-relevant outcomes. This modular execution flow allows Kumo to flexibly combine different workloads, schedulers, and attack strategies within a unified simulation framework.

\subsection{Workloads, Schedulers, and Attack Modeling}
\label{sec:modeling}

Kumo is designed to flexibly model serverless workloads, scheduling policies, and adversarial behavior within a unified simulation framework. These abstractions allow Kumo to express a wide range of security scenarios while maintaining a clear separation between workload generation, scheduling decisions, and attack logic.

\paragraph{Workloads}
Kumo provides multiple workload generators that capture common serverless invocation arrival patterns. Uniform workloads generate invocations by sampling tenants and functions evenly, and are primarily used for controlled stress testing and debugging. Poisson workloads model more realistic serverless traffic using exponentially distributed inter-arrival times, reflecting production environments. Kumo also supports bursty workloads that alternate between low- and high-intensity phases to capture flash crowds and workload spikes. All workload generators emit timestamped invocation arrivals without embedding any assumptions about placement or execution.

\paragraph{Schedulers}
Scheduling behavior is modeled through a unified scheduler abstraction that maps invocation requests to workers based on the current platform state. This abstraction allows different scheduling policies to be evaluated under identical workloads and platform configurations, isolating the security implications of scheduler behavior from other system components. In this work, we model several representative schedulers with distinct design objectives:
\begin{itemize}
    \item \textit{Random} selects a worker uniformly at random from the set of workers with sufficient available resources. It does not consider container locality, tenant identity, or prior placements, and serves as a baseline with no explicit isolation or performance optimization.
    \item \textit{DoubleDip}~\cite{shao2025bit} prioritizes spreading invocations across workers to reduce repeated co-location. It avoids placing invocations on workers that have recently hosted the same tenant when alternatives exist, and breaks ties by selecting less-loaded workers. This design captures schedulers that seek to improve isolation through placement diversity rather than strict partitioning.
    \item \textit{Helper}~\cite{zhao2024everywhere} prioritizes container reuse by favoring workers that already host warm containers for the invoked function. This locality-aware policy reduces cold-start overhead but increases the likelihood of repeated co-location, reflecting schedulers that optimize latency at the expense of isolation.
    \item \textit{OpenWhisk}~\cite{openwhisk} schedulers emphasize aggressive container reuse and opportunistic instance sharing. They preferentially place invocations on workers with existing containers for the same function, falling back to other workers only when necessary. While effective for performance, this behavior can amplify attacker--victim co-location.
\end{itemize}
These scheduler models are intentionally simplified to capture dominant placement behaviors relevant to security analysis, rather than to replicate proprietary production heuristics in full detail.

\paragraph{Attack Modeling}
To support security analysis, Kumo explicitly models attacker behavior at the workload level. Attacker modules wrap benign workloads and inject adversarial invocations that compete for shared platform resources. For co-location attacks, attackers issue invocations intended to increase the likelihood of sharing a worker with a victim tenant. For availability attacks, attackers inject invocations at configurable intensities to induce resource contention, queuing, and invocation drops. Attack parameters such as injection rate, arrival pattern, and service time are explicitly controlled, enabling systematic exploration of attacker capabilities.

Kumo allows experiments to designate specific tenants and functions as victims and tracks victim-specific metrics throughout execution. These metrics include co-location counts, time to first co-location, invocation drop rates, and tail latency. By explicitly labeling victims, Kumo enables direct measurement of security-relevant outcomes rather than relying on aggregate system metrics that may obscure attack impact.

Workloads, schedulers, and attacker models are fully configurable and composable. Any workload can be paired with any scheduler and attacker configuration without modifying the core engine, allowing Kumo to express a wide range of security scenarios using a small set of orthogonal building blocks.

\subsection{Experiment Configuration and Metrics}
\label{sec:config-metrics}

Kumo is driven by explicit experiment configuration files that fully specify platform parameters, workloads, schedulers, and attacker behavior. This design enables controlled, reproducible security experiments without modifying simulator code.

\paragraph{Configuration Parameters}
Each experiment configuration defines the serverless platform, workload characteristics, and security scenario under study. Platform parameters include the number of workers, per-worker CPU, memory, and storage resources, optional heterogeneity, container idle timeout, prewarming policy, and per-worker queue limits. Workloads are configurable as uniform, Poisson, or bursty arrival processes, with explicit control over arrival rates, batch sizes, and total invocation volume.

Scheduling policies are selected via a pluggable scheduler interface, and experiments may sweep across multiple schedulers and random seeds to capture stochastic effects. Attacker behavior is explicitly specified, including attacker tenant identity, attack intensity (expressed as attacker invocations per victim invocation), arrival pattern, and victim set. Service-time distributions are configurable (fixed or exponential), enabling controlled stress of system capacity.

Kumo supports parameter sweeps over schedulers, attacker intensity, queue length, cluster size, and other dimensions, allowing systematic exploration of security and performance tradeoffs.

\paragraph{Metrics Collection}
Kumo outputs a structured CSV trace summarizing both security- and performance-relevant outcomes for each experiment run. Reported metrics include:
(i) cold and warm start counts,
(ii) attacker--victim co-location count and time to first co-location,
(iii) total and victim-specific invocation arrivals and drops,
(iv) victim invocation drop rate,
(v) victim mean and tail (p95) latency,
and (vi) attacker drop rate.
Metrics are tracked per tenant, enabling direct measurement of victim impact rather than aggregate system behavior.
These metrics directly support the analyses in Case Studies A and B by quantifying isolation risk (co-location metrics) and availability degradation (drop rate and tail latency).

\paragraph{Tracing Support}
For debugging and validation, Kumo optionally records fine-grained execution traces capturing invocation arrivals, scheduling decisions, execution start and completion events, container reuse, and queuing behavior. Tracing is disabled by default and does not affect scheduling or execution semantics when enabled.

By combining explicit configuration, parameter sweeps, and security-focused metrics, Kumo enables reproducible and transparent evaluation of serverless security risks across a wide range of adversarial scenarios. These configuration and measurement capabilities are exercised in the case studies presented in Section~\ref{sec:evaluation}.

\section{Evaluation}
\label{sec:evaluation}

We evaluate Kumo through two complementary case studies that capture distinct classes of security risks in serverless platforms. 
Case Study~A examines \emph{co-location security}, where attackers attempt to achieve physical co-residency with a victim through scheduler placement decisions. 
Case Study~B focuses on \emph{availability (denial-of-service) attacks}, where attackers degrade victim performance through resource contention and queuing effects without exploiting scheduler logic.

Together, these case studies demonstrate Kumo’s ability to disentangle scheduler-driven isolation risks from system-level resource exhaustion vulnerabilities, and to support security analysis across multiple attack surfaces using a unified simulation framework.

\subsection{Case Study A: Attacker--Victim Co-location Security}
\label{sec:casestudy-colocation}

We first evaluate Kumo’s ability to analyze \emph{co-location security}, a prerequisite for many serverless side-channel and information leakage attacks. This case study examines how different serverless scheduling policies influence an attacker’s ability to become co-located with a victim tenant under identical workloads and platform configurations.

\subsubsection{Threat Model}
We consider a malicious tenant that repeatedly invokes an attack function in order to increase the likelihood of being scheduled on the same worker as a victim tenant. The attacker does not control the scheduler and cannot observe placement decisions directly. An attack is considered successful if an attacker invocation executes concurrently with a victim invocation on the same worker. This threat model is consistent with assumptions made by prior serverless co-location attacks.

\subsubsection{Experimental Setting}
We simulate a large-scale homogeneous serverless cluster consisting of 512 workers, each provisioned with identical CPU, memory, and storage resources. The platform hosts 200 benign tenants, each owning 20 functions, for a total of 4{,}000 distinct functions and 20{,}000 benign invocations. One benign tenant is designated as the victim, while a separate attacker tenant repeatedly invokes a dedicated attack function in an attempt to achieve co-location with the victim. All remaining tenants generate background traffic.

Invocation arrivals follow a Poisson process. Containers are subject to an idle timeout of 60 time units, and no pre-warming is enabled, reflecting a cost-efficient serverless configuration. The attacker injects invocations at a rate proportional to the victim’s workload, modeling an adversary that blends into normal system activity rather than overwhelming the platform.

We compare four schedulers—\textit{DoubleDip}, \textit{Random}, \textit{Helper}, and \textit{OpenWhisk}—under identical workload and platform conditions. Each experiment is repeated across 20 random seeds to capture stochastic effects, and all results are reported as means.

\begin{figure*}[ht!]
\centering
\includegraphics[width=0.9\linewidth]{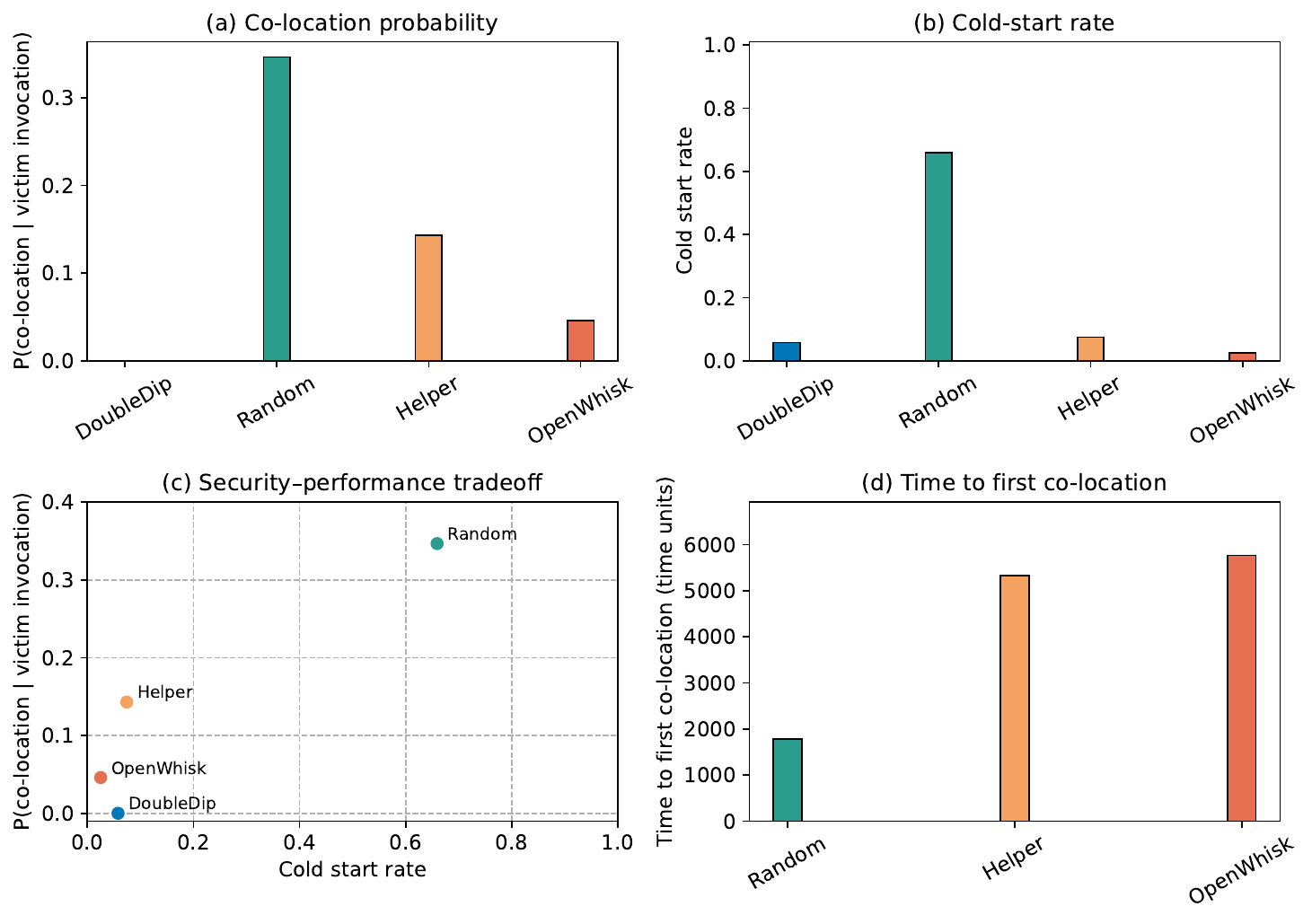}
\caption{Case Study A: Attacker--victim co-location under different schedulers in a large-scale serverless setting.}
\label{fig:expA}
\end{figure*}

Figure~\ref{fig:expA} summarizes the results of this case study.

\paragraph{Co-location Probability}
Figure~\ref{fig:expA}(a) reports the probability that an attacker invocation is co-located with the victim per victim invocation. Despite the large cluster size and high tenant diversity, co-location behavior differs substantially across schedulers. \textit{DoubleDip} effectively eliminates co-location, achieving zero probability across runs. In contrast, the \textit{Random} scheduler exhibits high co-location probability, while \textit{Helper} and \textit{OpenWhisk} fall between these extremes. These results illustrate how scheduler placement logic directly shapes co-location outcomes, even at large scale.

\paragraph{Cold-Start Overhead}
Figure~\ref{fig:expA}(b) shows the cold-start rate of the victim function under each scheduler. The \textit{Random} scheduler incurs a significantly higher cold-start rate, reflecting frequent container churn. In contrast, \textit{DoubleDip}, \textit{Helper}, and \textit{OpenWhisk} exhibit similarly low cold-start rates. This indicates that reduced co-location does not inherently require high cold-start overhead.

\paragraph{Security--Performance Tradeoff}
Figure~\ref{fig:expA}(c) visualizes the relationship between co-location probability and cold-start rate. \textit{DoubleDip} occupies a favorable region of the design space, combining low co-location probability with low cold-start overhead. Other schedulers exhibit less favorable tradeoffs, either exposing higher co-location risk or incurring increased cold-start cost. This visualization highlights how Kumo enables direct comparison of scheduler-induced security and performance characteristics.

\paragraph{Attack Feasibility}
Beyond steady-state co-location probability, we examine how quickly an attacker can succeed in practice. Figure~\ref{fig:expA}(d) reports the mean time to first attacker--victim co-location for schedulers that ever observe co-location. The \textit{Random} scheduler allows attackers to achieve co-location relatively quickly, while \textit{Helper} and \textit{OpenWhisk} delay initial co-location by several orders of magnitude in simulated time. Schedulers that never exhibit co-location are omitted from this plot. This time-based metric demonstrates that attack feasibility depends not only on probability but also on how rapidly co-location can occur.

This case study demonstrates that Kumo can capture systematic and persistent differences in co-location behavior across serverless schedulers, even in large-scale deployments with hundreds of workers and thousands of functions. The results show that scheduler placement policies strongly influence both the likelihood, which is consistent with existing literature~\cite{shao2025bit} and also reflects the timing of attacker--victim co-location, while their performance implications can be evaluated simultaneously within the same experimental framework.

\subsection{Case Study B: Denial-of-Service and Availability Degradation}
\label{sec:casestudy-dos}

\begin{figure*}[htb]
\centering
\includegraphics[width=0.9\linewidth]{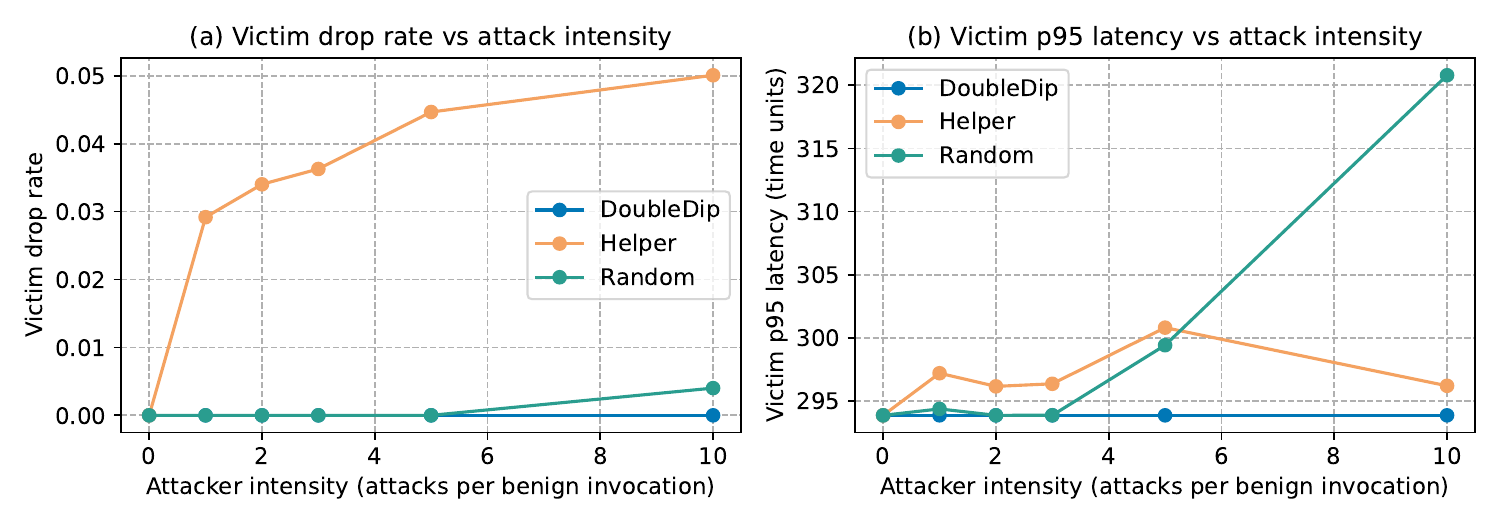}
\caption{Case Study B: Victim availability and tail latency under increasing attacker intensity.}
\label{fig:dos-intensity}
\end{figure*}

\begin{figure*}[htb]
\centering
\includegraphics[width=0.9\linewidth]{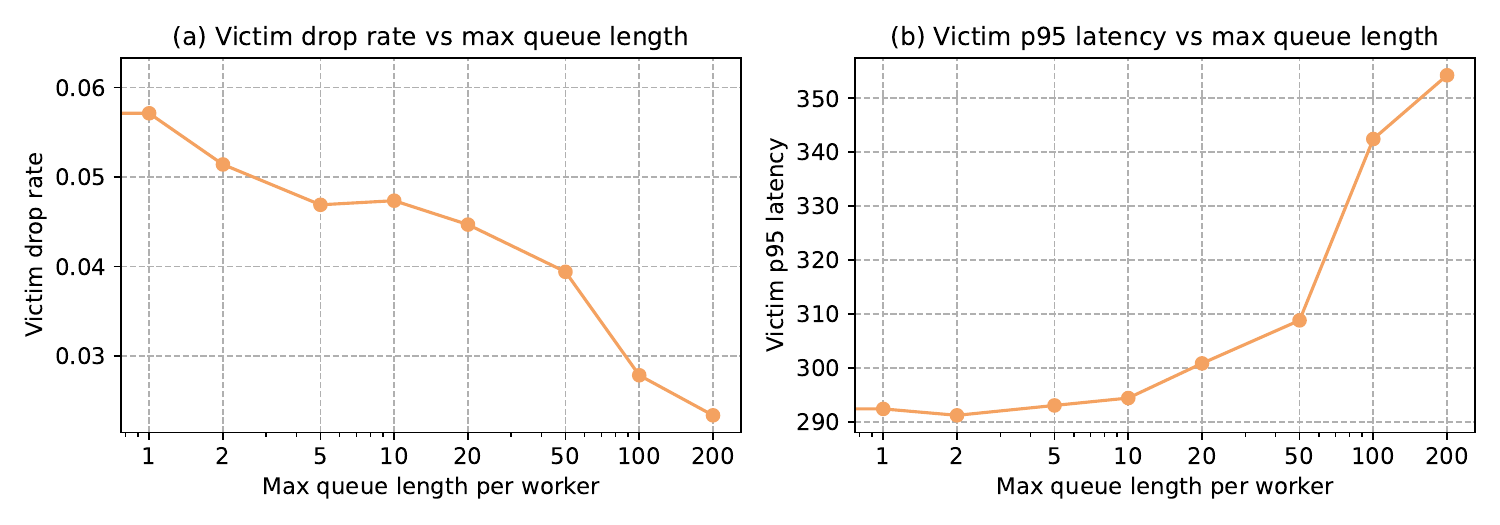}
\caption{Case Study B: Impact of per-worker queue length on availability and tail latency.}
\label{fig:dos-queue}
\end{figure*}

\begin{figure*}[htb]
\centering
\includegraphics[width=0.9\linewidth]{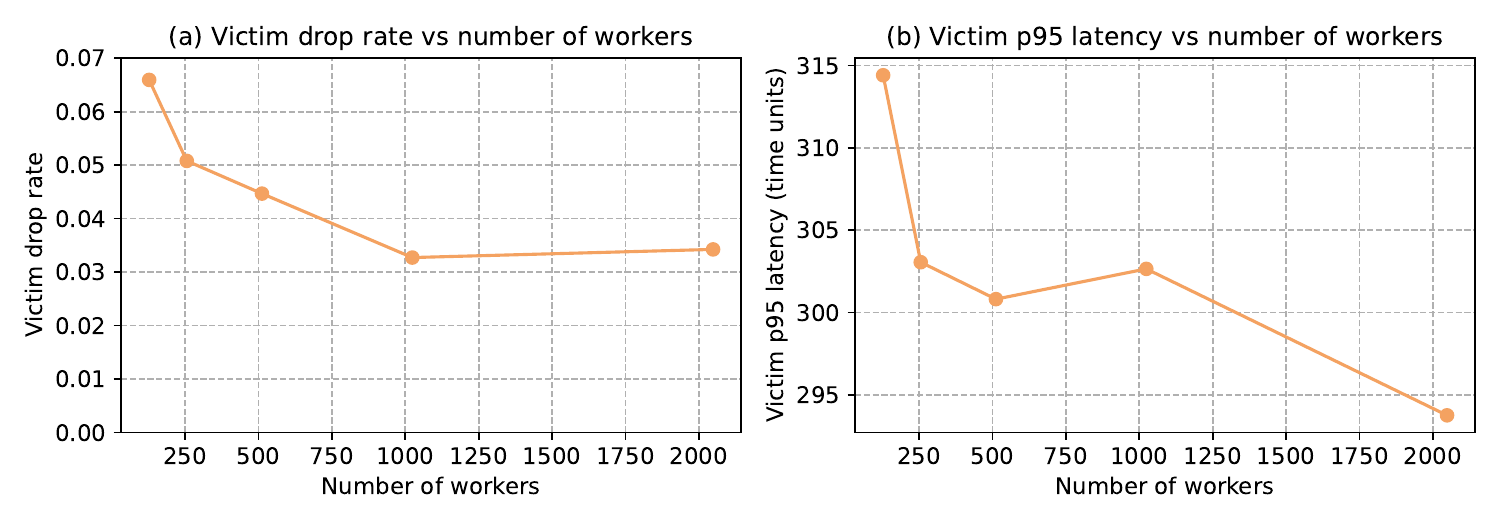}
\caption{Case Study B: Effect of cluster scale on victim availability and tail latency under fixed attacker intensity.}
\label{fig:dos-scale}
\end{figure*}

We next demonstrate Kumo’s ability to analyze \emph{availability attacks}, focusing on denial-of-service (DoS) behavior in serverless platforms. Unlike co-location attacks, which exploit scheduler placement decisions, DoS attacks aim to degrade victim availability and latency through sustained resource contention and queuing effects. This case study examines how attacker intensity, queueing policies, and cluster scale influence DoS impact under realistic multi-tenant serverless workloads.

\paragraph{Threat Model}
We consider an attacker tenant that issues a large volume of legitimate function invocations to contend for shared worker resources. The attacker does not crash workers, exploit implementation bugs, or generate network-level floods. Instead, the attack relies solely on invocation-level pressure to exhaust execution capacity and queue space. An attack is considered successful if it increases victim invocation drop rates or significantly inflates tail latency.

\paragraph{Experimental Setting}
We evaluate DoS behavior using three controlled experiments that vary attacker intensity, queuing policy, and cluster scale. Unless otherwise stated, we simulate a homogeneous serverless cluster with 512 workers, each provisioned with identical CPU, memory, and storage resources. The platform hosts 200 benign tenants, each owning 20 functions, and 20{,}000 benign invocations in total, generating background traffic representative of large multi-tenant deployments.

Workloads follow a Poisson arrival process, with arrival rates tuned such that the system operates near its capacity threshold even without an attacker. Service times follow an exponential distribution with a mean of 100 time units, capturing realistic execution variability. Containers are subject to a fixed idle timeout, and pre-warming is disabled. Each experiment is repeated across 20 random seeds, and results are reported as means.

Specifically, (i)~to study attacker intensity, we sweep the scheduler and attacker injection rate while holding cluster size and queue limits fixed; (ii)~to study queuing effects, we sweep the maximum per-worker queue length under a fixed attacker intensity and Helper scheduler; and (iii)~to study scalability, we vary the number of workers while keeping workload characteristics constant and Helper scheduler. These experiments correspond to Figures~\ref{fig:dos-intensity}, \ref{fig:dos-queue}, and \ref{fig:dos-scale}, respectively.

\paragraph{Impact of Attacker Intensity}
Figure~\ref{fig:dos-intensity} sweeps attacker intensity from $0$ to $10$ attacks intensities and measures victim drop rate and p95 latency. Two observations stand out. First, DoS impact increases with attacker intensity as expected: higher adversarial load pushes the system closer to saturation, which increases both the probability of queue overflow (drops) and the waiting time of successfully served invocations (tail latency). Second, the scheduler policy does affect DoS outcomes in this near-capacity regime. In particular, the Helper scheduler exhibits noticeably higher victim drop rates as intensity grows, while DoubleDip remains zero drop rate and nearly flat p95 latency across the sweep. The Random scheduler behaves in-between: it largely avoids drops at low-to-moderate intensity, but its tail latency rises and drops appear at the highest intensity. These differences are consistent with the intuition that schedulers can either concentrate contention (creating hotspots that overflow queues) or spread load to avoid per-worker collapse, even when the total offered load is the same. Kumo makes this observation explicit by modeling resource contention and backpressure directly.

\paragraph{Queue Length Tradeoffs}
Figure~\ref{fig:dos-queue} sweeps the per-worker maximum queue length under a fixed attacker intensity (5.0). Increasing queue capacity reduces victim drop rate because fewer requests are rejected when workers are temporarily saturated. However, this mitigation comes with a clear tail-latency cost: as queue length grows, more requests wait longer before service, increasing victim p95 latency. The curve shows a visible ``knee'' (roughly around queue lengths in the tens to low hundreds), illustrating that modest queueing can reduce drops without severely harming p95, while aggressive queueing increasingly converts drops into long tail latency. By modeling queue limits and execution delays, Kumo enables systematic exploration of this tradeoff under adversarial load.

\paragraph{Effect of Cluster Scale}
Figure~\ref{fig:dos-scale} sweeps cluster size from $128$ to $2048$ workers under a fixed attacker intensity (5.0) and queue limit (50). Increasing the number of workers improves victim availability: drop rate decreases as additional capacity absorbs load, and tail latency trends downward overall. However, the improvement is not perfectly linear and exhibits diminishing returns, suggesting that beyond a certain point the remaining drops and tail latency are dominated by transient bursts, queueing effects, and stochastic imbalance rather than raw capacity alone. This behavior reflects a more realistic operating regime in which capacity expansion mitigates, but does not fully eliminate, contention-induced degradation. By allowing cluster size to be swept independently of workload and attacker behavior, Kumo enables precise analysis of how horizontal scaling shifts the availability–latency tradeoff under adversarial load.

This case study demonstrates that Kumo can model availability degradation driven by resource contention, queuing, and capacity limits, and can expose how design choices shape DoS behavior. In our scaled setting, DoS impact depends not only on attacker intensity and queue sizing, but also on scheduler behavior: some policies amplify hotspots and drops under overload, while others spread load and stabilize victim tail latency. Together with Case Study~A, these results highlight why security analysis for serverless platforms must jointly consider scheduler policy, contention dynamics, and queueing limits.

\subsection{Lessons Learned}
\label{sec:lessons}

Our evaluation highlights several lessons about analyzing security risks in serverless platforms.

First, \emph{scheduler behavior can act as a security-relevant control surface}. Case Study~A shows that different scheduling policies, when evaluated under identical workloads and platform configurations, can lead to substantial differences in attacker--victim co-location probability and time to first co-location. This demonstrates that placement decisions alone can meaningfully influence isolation risk, and that schedulers should be treated as configurable security mechanisms rather than fixed platform components.

Second, \emph{availability degradation emerges from the interaction between scheduling, queuing, and capacity}. As illustrated in Case Study~B, denial-of-service behavior depends strongly on attacker intensity, service time, queue limits, and cluster scale. While schedulers may influence how contention manifests (e.g., through load spreading versus hotspot formation), availability outcomes are ultimately shaped by system-level resource dynamics once the platform approaches saturation.

Finally, \emph{security and performance tradeoffs are inherently multi-dimensional}. Improving isolation may introduce moderate performance overheads, while mitigating availability attacks often shifts failures from drops to increased tail latency through deeper queuing or requires additional capacity through scaling. These tradeoffs cannot be captured by a single metric alone or configuration choice, underscoring the need for analysis tools that jointly model scheduler behavior, resource contention, and workload dynamics.

Together, these lessons reinforce the value of Kumo as a simulation framework for disentangling scheduler-driven security risks from broader resource exhaustion effects, enabling systematic exploration of security--performance tradeoffs in serverless platforms.

\subsection{Simulation Performance}
\label{sec:sim-performance}

We report the execution time of our simulations to demonstrate the practicality of Kumo for large-scale security analysis. All experiments were run on a single commodity desktop machine, using a single core and no parallelization.

Case Study~A consists of one experiment with $4$ schedulers and $20$ random seeds, corresponding to $80$ simulation runs in total. Case Study~B consists of three DoS experiments:
\begin{itemize}
    \item \textbf{Attack intensity sweep (B1):} With $3$ schedulers, $6$ attacker intensities, and $20$ random seeds, resulting in $360$ runs.
    \item \textbf{Queue length sweep (B2):} With $9$ per-worker queue length settings (including queue length $=0$) and $20$ random seeds, resulting in $180$ runs.
    \item \textbf{Number of workers sweep (B3):} With $5$ cluster sizes and $20$ random seeds, resulting in $100$ runs.
\end{itemize}

The runtime of each experiment is summarized in Table~\ref{tab:runtime_summary}.

\begin{table}[t]
\centering
\caption{Runtime summary for Case Studies A and B.}
\label{tab:runtime_summary}
\begin{tabular}{l c c c}
\hline
\textbf{Experiment} & \textbf{Total Time (s)} & \textbf{Total Runs} & \textbf{Time / Run (s)} \\
\hline
Case Study~A  & 394.1 & 80  & 4.93 \\
Case Study~B1 & 777.9 & 360 & 2.16 \\
Case Study~B2 & 353.8 & 180 & 1.97 \\
Case Study~B3 & 275.7 & 100 & 2.76 \\
\hline
\end{tabular}
\end{table}

These results show that Kumo can simulate tens of thousands of invocations across hundreds of workers and tenants within minutes on a single machine. This performance makes Kumo suitable for iterative security analysis, large parameter sweeps, and comparative evaluation of scheduling and defense mechanisms.

\section{Discussion}
\label{sec:discussion}


\subsection{Scheduler-Driven vs.\ System-Level Security}

A key insight from our evaluation is the clear separation between scheduler-driven isolation risks and system-level availability vulnerabilities. Case Study~A shows that scheduler choice is a first-order security mechanism for co-location attacks: under identical workloads and platform configurations, different schedulers lead to orders-of-magnitude differences in attacker--victim co-location probability and attack feasibility. In contrast, Case Study~B demonstrates that DoS behavior is largely insensitive to scheduler design once resource contention dominates, and is instead governed by factors such as service time, queuing policy, and cluster capacity.

This distinction suggests that no single mechanism can address all serverless security risks. Scheduler design plays a critical role in mitigating isolation attacks, but availability attacks require system-level defenses that go beyond placement decisions. Kumo enables these distinctions to be studied explicitly within a unified framework.

\subsection{Implications for Serverless Platform Design}

Our results highlight several implications for serverless platform operators. First, schedulers should be viewed not only as performance optimizers but also as security mechanisms. Policies that reduce co-location risk can substantially improve isolation with modest performance overhead. Second, defenses against DoS attacks cannot rely solely on scheduling. Queue management, admission control, and capacity provisioning all play a central role in determining availability under attack.

Importantly, several commonly used mitigation strategies involve tradeoffs. Increasing queue capacity reduces invocation drops but can severely degrade tail latency, while horizontal scaling improves availability at significant infrastructure cost. These tradeoffs underscore the need for principled analysis tools that expose the security consequences of design choices before deployment.

\subsection{Limitations}

Kumo is intentionally scoped to balance fidelity and tractability. It does not model low-level microarchitectural side channels, network-level DoS, or platform-specific implementation details such as proprietary placement heuristics. Instead, Kumo focuses on system-level behavior arising from shared execution environments, container reuse, and resource contention.

While this abstraction limits Kumo’s ability to study certain classes of attacks, it enables scalable and reproducible analysis of dominant security mechanisms in serverless platforms. Future work could incorporate additional layers of fidelity where needed, without altering Kumo’s core architecture.

\subsection{Reproducibility and Availability}

To support reproducibility and open-science practices, we release the complete Kumo artifact as open-source software. The artifact has been independently evaluated by the CCGRID Artifact Evaluation Committee and awarded the \textbf{Results Reproduced (ROR-R)} badge. 

Additional details, including installation instructions and reproduction steps, are provided in Appendix.

\subsection{Future Directions}

Kumo opens several avenues for future research. One direction is integrating cost models to study the economic impact of security defenses, such as over-provisioning or aggressive isolation. Another is exploring scheduler synthesis, where security constraints are incorporated directly into scheduler design and evaluated using Kumo’s simulation framework. 


\section{Related Work}
\label{sec:related-work}
In contrast to prior work that focuses on either serverless performance optimization or empirical security attacks, Kumo bridges these domains by providing a security-focused simulation framework. We organize related work into four areas.

\subsection{Serverless Platforms and Scheduling}
Prior work has extensively studied the design and optimization of serverless platforms, focusing on scheduling, performance, and cost efficiency. Several studies propose scheduling policies to reduce cold starts, improve latency, or optimize resource utilization in serverless systems. Production platforms such as AWS Lambda, Azure Functions, and OpenWhisk employ sophisticated placement and reuse heuristics, though their exact implementations are largely opaque~\cite{awslambda,azure,openwhisk}. While these works provide important insights into serverless performance, they generally do not consider scheduler behavior as a security mechanism or analyze attacker-driven workloads.

\subsection{Serverless Simulation and Modeling}
A number of simulators and analytical models have been proposed for serverless computing, primarily targeting performance evaluation, cost modeling, and capacity planning~\cite{mahmoudi2021simfaas,raith2023faas,mampage2023cloudsimsc}. These tools typically model invocation arrival processes, execution latency, and resource consumption, and are useful for understanding scalability and efficiency tradeoffs. However, existing simulators rarely include explicit attacker modeling, security-oriented metrics, or mechanisms to study isolation and availability under adversarial conditions. Kumo complements this line of work by prioritizing security analysis and explicitly modeling attackers, victims, and scheduler-induced isolation effects.

\subsection{Co-location and Side-Channel Attacks in the Cloud}
Co-location attacks have been widely studied in virtualized and multi-tenant cloud environments, where attackers attempt to place workloads on the same physical host as a victim to exploit side channels. Recent work has extended these attacks to serverless platforms, demonstrating that scheduler placement decisions and container reuse can enable information leakage even in short-lived execution environments~\cite{fang2023heteroscore,fang2022repttack}. These studies typically rely on empirical experiments on production platforms or small-scale testbeds. Kumo provides a complementary approach by enabling controlled, reproducible exploration of co-location risk across different scheduling policies without requiring access to proprietary systems.

\subsection{Denial-of-Service and Resource Exhaustion Attacks}
DoS attacks through resource exhaustion have long been studied in cloud and distributed systems. Prior work has examined admission control, queuing policies, and over-provisioning as defenses against overload and performance collapse~\cite{cho2020overload,park2024topfull}. In serverless settings, recent studies have highlighted how bursty workloads and unbounded scaling assumptions can lead to unexpected availability degradation~\cite{suresh2020ensure,qiu2021function}. Unlike traditional network-level DDoS attacks, these behaviors arise from legitimate invocation patterns. Kumo models this class of attacks by explicitly capturing resource contention, queuing, and invocation drops, enabling systematic study of availability tradeoffs under adversarial load.

\section{Conclusion}
\label{sec:conclusion}

In this paper, we presented Kumo, a security-focused simulator for serverless platforms that enables principled analysis of risks arising from scheduling and resource sharing decisions. Kumo is designed to support controlled, reproducible experiments with explicit modeling of workloads, schedulers, attackers, and victims, bridging a gap between performance-oriented simulators and empirical security studies on production systems.

Through two complementary case studies, we demonstrated how Kumo can be used to analyze distinct classes of serverless security risks. Our co-location study showed that scheduler choice is a first-order security mechanism, with different policies inducing orders-of-magnitude differences in attacker--victim co-location probability and attack feasibility. In contrast, our DoS study revealed that availability degradation is largely governed by system-level factors such as resource contention, queuing behavior, and cluster capacity, and is comparatively insensitive to scheduler design once contention dominates. Together, these results highlight the importance of distinguishing scheduler-driven isolation risks from broader resource exhaustion vulnerabilities.

Kumo provides a flexible foundation for exploring these tradeoffs without requiring access to proprietary platforms or costly real-world experimentation. By enabling systematic, security-aware evaluation of serverless design choices, Kumo complements existing empirical work and opens new avenues for research on secure and resilient serverless computing.

\section{Acknowledgment}
The work in this paper is partially supported by the National Science Foundation (NSF) grants CNS-2155002 and 2311888.

\bibliographystyle{IEEEtran}
\bibliography{ref.bib}

@article{shao2025bit,
  title={Bit of a Close Talker: A Practical Guide to Serverless Cloud Co-Location Attacks},
  author={Shao, Wei and Nazari, Najmeh and Omidi, Behnam and Rafatirad, Setareh and Homayoun, Houman and Khasawneh, Khaled N and Fang, Chongzhou},
  journal={arXiv preprint arXiv:2512.10361},
  year={2025}
}

@article{li2022serverless,
  title={Serverless computing: state-of-the-art, challenges and opportunities},
  author={Li, Yongkang and Lin, Yanying and Wang, Yang and Ye, Kejiang and Xu, Chengzhong},
  journal={IEEE Transactions on Services Computing},
  volume={16},
  number={2},
  pages={1522--1539},
  year={2022},
  publisher={IEEE}
}

@article{shafiei2022serverless,
  title={Serverless computing: a survey of opportunities, challenges, and applications},
  author={Shafiei, Hossein and Khonsari, Ahmad and Mousavi, Payam},
  journal={ACM Computing Surveys},
  volume={54},
  number={11s},
  pages={1--32},
  year={2022},
  publisher={ACM New York, NY}
}

@article{ghorbian2024survey,
  title={A survey on the scheduling mechanisms in serverless computing: a taxonomy, challenges, and trends},
  author={Ghorbian, Mohsen and Ghobaei-Arani, Mostafa and Esmaeili, Leila},
  journal={Cluster Computing},
  volume={27},
  number={5},
  pages={5571--5610},
  year={2024},
  publisher={Springer}
}

@inproceedings{fang2022repttack,
  title={Repttack: Exploiting Cloud Schedulers to Guide Co-Location Attacks},
  author={Fang, Chongzhou and Wang, Han and Nazari, Najmeh and Omidi, Behnam and Sasan, Avesta and Khasawneh, Khaled N and Rafatirad, Setareh and Homayoun, Houman},
  booktitle={Proceedings of the Network and Distributed Systems Security (NDSS) Symposium},
  year={2022}
}

@article{marin2022serverless,
  title={Serverless computing: a security perspective},
  author={Marin, Eduard and Perino, Diego and Di Pietro, Roberto},
  journal={Journal of Cloud Computing},
  volume={11},
  number={1},
  pages={69},
  year={2022},
  publisher={Springer}
}

@inproceedings{zhao2024everywhere,
  title={Everywhere all at once: Co-location attacks on public cloud faas},
  author={Zhao, Zirui Neil and Morrison, Adam and Fletcher, Christopher W and Torrellas, Josep},
  booktitle={Proceedings of the 29th ACM International Conference on Architectural Support for Programming Languages and Operating Systems, Volume 1},
  pages={133--149},
  year={2024}
}

@article{mahmoudi2021simfaas,
  title={SimFaaS: A performance simulator for serverless computing platforms},
  author={Mahmoudi, Nima and Khazaei, Hamzeh},
  journal={arXiv preprint arXiv:2102.08904},
  year={2021}
}

@article{raith2023faas,
  title={faas-sim: A trace-driven simulation framework for serverless edge computing platforms},
  author={Raith, Philipp and Rausch, Thomas and Furutanpey, Alireza and Dustdar, Schahram},
  journal={Software: Practice and Experience},
  volume={53},
  number={12},
  pages={2327--2361},
  year={2023},
  publisher={Wiley Online Library}
}

@inproceedings{mampage2023cloudsimsc,
  title={CloudSimSC: A toolkit for modeling and simulation of serverless computing environments},
  author={Mampage, Anupama and Buyya, Rajkumar},
  booktitle={2023 IEEE International Conference on High Performance Computing \& Communications, Data Science \& Systems, Smart City \& Dependability in Sensor, Cloud \& Big Data Systems \& Application (HPCC/DSS/SmartCity/DependSys)},
  pages={550--557},
  year={2023},
  organization={IEEE}
}

@inproceedings{gruss2016flush+,
  title={Flush+ flush: A fast and stealthy cache attack},
  author={Gruss, Daniel and Maurice, Cl{\'e}mentine and Wagner, Klaus and Mangard, Stefan},
  booktitle={International Conference on Detection of Intrusions and Malware, and Vulnerability Assessment},
  pages={279--299},
  year={2016},
  organization={Springer}
}

@article{kocher2020spectre,
  title={Spectre attacks: Exploiting speculative execution},
  author={Kocher, Paul and Horn, Jann and Fogh, Anders and Genkin, Daniel and Gruss, Daniel and Haas, Werner and Hamburg, Mike and Lipp, Moritz and Mangard, Stefan and Prescher, Thomas and others},
  journal={Communications of the ACM},
  volume={63},
  number={7},
  pages={93--101},
  year={2020},
  publisher={ACM New York, NY, USA}
}

@article{lipp2020meltdown,
  title={Meltdown: Reading kernel memory from user space},
  author={Lipp, Moritz and Schwarz, Michael and Gruss, Daniel and Prescher, Thomas and Haas, Werner and Horn, Jann and Mangard, Stefan and Kocher, Paul and Genkin, Daniel and Yarom, Yuval and others},
  journal={Communications of the ACM},
  volume={63},
  number={6},
  pages={46--56},
  year={2020},
  publisher={ACM New York, NY, USA}
}

@inproceedings{yarom2014flush+,
  title={$\{$FLUSH+ RELOAD$\}$: A high resolution, low noise, l3 cache $\{$Side-Channel$\}$ attack},
  author={Yarom, Yuval and Falkner, Katrina},
  booktitle={23rd USENIX security symposium (USENIX security 14)},
  pages={719--732},
  year={2014}
}

@inproceedings{ishakian2010colocation,
  title={Colocation as a service: Strategic and operational services for cloud colocation},
  author={Ishakian, Vatche and Sweha, Raymond and Londono, Jorge and Bestavros, Azer},
  booktitle={2010 Ninth IEEE International Symposium on Network Computing and Applications},
  pages={76--83},
  year={2010},
  organization={IEEE}
}

@inproceedings{azar2014co,
  title={Co-location-resistant clouds},
  author={Azar, Yossi and Kamara, Seny and Menache, Ishai and Raykova, Mariana and Shepard, Bruce},
  booktitle={Proceedings of the 6th Edition of the ACM Workshop on Cloud Computing Security},
  pages={9--20},
  year={2014}
}

@inproceedings{fang2023heteroscore,
  title={Heteroscore: Evaluating and mitigating cloud security threats brought by heterogeneity},
  author={Fang, Chongzhou and Nazari, Najmeh and Omidi, Behnam and Wang, Han and Puri, Aditya and Arora, Manish and Rafatirad, Setareh and Homayoun, Houman and Khasawneh, Khaled N},
  booktitle={The Network and Distributed System Security Symposium (NDSS)},
  year={2023}
}

@article{nabi2016availability,
  title={Availability in the cloud: State of the art},
  author={Nabi, Mina and Toeroe, Maria and Khendek, Ferhat},
  journal={Journal of Network and Computer Applications},
  volume={60},
  pages={54--67},
  year={2016},
  publisher={Elsevier}
}

@article{sharifi2012availability,
  title={Availability challenge of cloud system under DDOS attack},
  author={Sharifi, Aboosaleh Mohammad and Amirgholipour, Saeed K and Alirezanejad, Mehdi and Aski, Baharak Shakeri and Ghiami, Mohammad},
  journal={Indian Journal of Science and Technology},
  volume={5},
  number={6},
  pages={2933--7},
  year={2012},
  publisher={Indian Society for Education and Environment, 23(new) Neelkamal Apt, 3 d~…}
}

@misc{openwhisk,
  author = {OpenWhisk},
  title = {{OpenWhisk Documentation}},
  howpublished = "\url{https://openwhisk.apache.org/documentation.html}",
  note = "[Online; accessed Jun. 2024]"
}

@misc{awslambda,
  author = {Amazon},
  title = {{AWS Lambda}},
  howpublished = "\url{https://aws.amazon.com/lambda/}",
  note = "[Online; accessed Jun. 2024]"
}

@misc{azure,
  title={{Microsoft Azure}},
  author={Microsoft Azure},
  year        = {2022},
  howpublished = "\url{https://azure.microsoft.com/en-us/}",
  note = "[Online; accessed Apr. 2022]"
}

@inproceedings{cho2020overload,
  title={Overload control for $\{$$\mu$s-scale$\}$$\{$RPCs$\}$ with breakwater},
  author={Cho, Inho and Saeed, Ahmed and Fried, Joshua and Park, Seo Jin and Alizadeh, Mohammad and Belay, Adam},
  booktitle={14th USENIX Symposium on Operating Systems Design and Implementation (OSDI 20)},
  pages={299--314},
  year={2020}
}

@inproceedings{park2024topfull,
  title={TopFull: An Adaptive Top-Down Overload Control for SLO-Oriented Microservices},
  author={Park, Jinwoo and Park, Jaehyeong and Jung, Youngmok and Lim, Hwijoon and Yeo, Hyunho and Han, Dongsu},
  booktitle={Proceedings of the ACM SIGCOMM 2024 Conference},
  pages={876--890},
  year={2024}
}

@inproceedings{suresh2020ensure,
  title={Ensure: Efficient scheduling and autonomous resource management in serverless environments},
  author={Suresh, Amoghavarsha and Somashekar, Gagan and Varadarajan, Anandh and Kakarla, Veerendra Ramesh and Upadhyay, Hima and Gandhi, Anshul},
  booktitle={2020 IEEE International Conference on Autonomic Computing and Self-Organizing Systems (ACSOS)},
  pages={1--10},
  year={2020},
  organization={IEEE}
}

@inproceedings{qiu2021function,
  title={Is function-as-a-service a good fit for latency-critical services?},
  author={Qiu, Haoran and Jha, Saurabh and Banerjee, Subho S and Patke, Archit and Wang, Chen and Hubertus, Franke and Kalbarczyk, Zbigniew T and Iyer, Ravishankar K},
  booktitle={Proceedings of the Seventh International Workshop on Serverless Computing (WoSC7) 2021},
  pages={1--8},
  year={2021}
}

@article{nazari2023adversarial,
  title={Adversarial attacks against machine learning-based resource provisioning systems},
  author={Nazari, Najmeh and Makrani, Hosein Mohammadi and Fang, Chongzhou and Omidi, Behnam and Rafatirad, Setareh and Sayadi, Hossein and Khasawneh, Khaled N and Homayoun, Houman},
  journal={IEEE Micro},
  volume={43},
  number={5},
  pages={35--44},
  year={2023},
  publisher={IEEE}
}

@inproceedings{fang2023gotcha,
  title={Gotcha! i know what you are doing on the fpga cloud: Fingerprinting co-located cloud fpga accelerators via measuring communication links},
  author={Fang, Chongzhou and Miao, Ning and Wang, Han and Zhou, Jiacheng and Sheaves, Tyler and Emmert, John M and Sasan, Avesta and Homayoun, Houman},
  booktitle={Proceedings of the 2023 ACM SIGSAC Conference on Computer and Communications Security},
  pages={2024--2037},
  year={2023}
}

@article{fang2025assessing,
  title={Assessing and Mitigating Heterogeneity-Driven Security Threats in the Cloud},
  author={Fang, Chongzhou and Nazari, Najmeh and Omidi, Behnam and Wang, Han and Puri, Aditya and Arora, Manish and Rafatirad, Setareh and Homayoun, Houman and Khasawneh, Khaled},
  journal={ACM Transactions on Internet Technology},
  volume={25},
  number={4},
  pages={1--31},
  year={2025},
  publisher={ACM New York, NY}
}

\appendix

\subsection{Artifact Description}

\subsubsection{Overview}

This artifact contains the complete implementation of \textit{Kumo}, a discrete-event simulator for analyzing security risks in serverless platforms, along with all configuration files and plotting scripts required to reproduce the results presented in the paper.

The artifact enables reproduction of:

\begin{itemize}
    \item Section IV-A: Case Study A (Attacker--Victim Co-location)
    \item Section IV-B: Case Study B (Denial-of-Service Experiments)
    \item Figures 2--5
    \item Table I (runtime measurements)
\end{itemize}

The archived artifact is available at:

\begin{itemize}
    \item DOI: \url{https://doi.org/10.5281/zenodo.18635971}
    \item Development repository: \newline\url{https://github.com/YoungKameSennin/kumo}
\end{itemize}

\subsubsection{Artifact Contents}

The artifact includes:

\begin{itemize}
    \item \texttt{src/} --- C++17 implementation of the Kumo simulator
    \item \texttt{configs/} --- Experiment configurations for Case Studies A, B1, B2, and B3
    \item \texttt{plot/} --- Python scripts for generating all paper figures
    \item \texttt{Makefile} --- Build system for the simulator
    \item \texttt{artifacts\_reproduce.sh} --- One-command reproduction script
    \item \texttt{requirements.txt} --- Python plotting dependencies
\end{itemize}

The simulator is implemented in C++17 and uses Python for data analysis and plotting.

\subsubsection{System Requirements}

The artifact runs on commodity hardware.

\textbf{Hardware requirements}

\begin{itemize}
    \item Standard laptop or desktop
    \item Single CPU core sufficient
    \item No GPU required
\end{itemize}

\textbf{Software requirements}

\begin{itemize}
    \item Linux or macOS
    \item \texttt{g++} with C++17 support (GCC $\geq$ 7 recommended)
    \item Python 3.9+
    \item Python packages: \texttt{numpy}, \texttt{pandas}, \texttt{matplotlib}
\end{itemize}

Install Python dependencies:

\begin{verbatim}
    pip install -r requirements.txt
\end{verbatim}

\subsubsection{Build Instructions}

Compile the simulator:

\begin{verbatim}
    make
\end{verbatim}

This produces the binary \texttt{kumo\_experiment}.

\subsubsection{Reproducing Experimental Results}

To reproduce all results:

\begin{verbatim}
    make reproduce
\end{verbatim}

The reproduction script performs the following steps:

\begin{enumerate}
    \item Builds the simulator if necessary.
    \item Executes all experiment configurations in \texttt{configs/}.
    \item Generates result CSV files in \texttt{results/}.
    \item Generates figures in \texttt{figs/}.
    \item Performs sanity checks to confirm expected outputs exist.
\end{enumerate}

\subsubsection{Expected Outputs}

After successful reproduction:

\begin{itemize}
    \item CSV result files:
    \begin{itemize}
        \item \texttt{case\_study\_A\_result.csv}
        \item \texttt{case\_study\_B1\_result.csv}
        \item \texttt{case\_study\_B2\_result.csv}
        \item \texttt{case\_study\_B3\_result.csv}
    \end{itemize}
    \item Figures corresponding to:
    \begin{itemize}
        \item Figure 2 (Co-location analysis)
        \item Figure 3 (Attack intensity sweep)
        \item Figure 4 (Queue length sweep)
        \item Figure 5 (Cluster size sweep)
    \end{itemize}
\end{itemize}

\subsubsection{Estimated Runtime}

On a commodity desktop:

\begin{itemize}
    \item Build time: $<$ 10 seconds
    \item Case Study A: $\sim$ 5 seconds per run
    \item Case Study B experiments: $\sim$ 2--3 seconds per run
    \item Full reproduction: typically 30--60 minutes
\end{itemize}

These runtimes are consistent with Table I in the paper.

\subsubsection{Mapping to Paper Results}

\begin{center}
\begin{tabular}{ll}
\toprule
\textbf{Paper Result} & \textbf{Artifact Component} \\
\midrule
Figure 2(a--d) & \texttt{configs/case\_study\_A.cfg} \\
Figure 3 & \texttt{configs/case\_study\_B1.cfg} \\
Figure 4 & \texttt{configs/case\_study\_B2.cfg} \\
Figure 5 & \texttt{configs/case\_study\_B3.cfg} \\
Table I & Execution logs and runtime measurements \\
\bottomrule
\end{tabular}
\end{center}

\subsubsection{Verification Guidelines}

Readers may verify the following behaviors:

\begin{itemize}
    \item The DoubleDip scheduler exhibits near-zero co-location probability.
    \item Increasing attacker intensity increases victim drop rate.
    \item Larger queue limits reduce drops but increase tail latency.
    \item Increasing cluster size reduces drop rate and latency.
\end{itemize}

Exact numerical equality is not required; reproduction focuses on behavioral consistency and trends.

\subsubsection{Reproduction Steps}

The artifact includes fixed configuration files and deterministic random seeds to ensure reproducible execution.

To reproduce the results reported in the paper:

\begin{enumerate}
    \item Build the simulator:
\begin{verbatim}
make
\end{verbatim}

    \item Install Python dependencies:
\begin{verbatim}
pip install -r requirements.txt
\end{verbatim}

    \item Execute all experiments and generate plots:
\begin{verbatim}
make reproduce
\end{verbatim}

    \item After completion, CSV outputs will appear in \texttt{results/} and generated figures will appear in \texttt{figs/}.

    \item Compare the generated figures with Figures 2--5 in the paper to verify behavioral consistency.
\end{enumerate}

\subsubsection{License}

The artifact is released under the MIT License.

\subsubsection{Summary}

This artifact provides a complete, self-contained implementation of Kumo and enables reproduction of all experimental results using commodity hardware. The artifact has been independently verified by the CCGRID Artifact Evaluation Committee and awarded the \textbf{Results Reproduced (ROR-R)} badge.

\end{document}